\documentclass[useAMS,usenatbib]{mn2e}
\usepackage{times}

\usepackage{natbib,amsmath,amssymb}
\usepackage{graphicx}%,massymb}

\newcommand{\numnuv}{102385}
\newcommand{\nodupg}{98629}
\newcommand{\numgk}{39437}
\newcommand{\nuv}{$NUV$}
\newcommand{\fuv}{$FUV$}

\newcommand{\sqdeg}{deg$^2$}
\newcommand\ion[2]{#1$\;${\small\rmfamily\@Roman{#2}}\relax}%

\newcommand\nodata{{...}}%
\newcommand\mnras{{MNRAS}}%
\newcommand\apj{{ApJ}}%
\newcommand\apjl{{ApJ}}%
\newcommand\apjs{{ApJS}}%
\newcommand\araa{{ARA\&A}}%
\newcommand\aj{{AJ}}%
\newcommand\pasp{{PASP}}%

\begin{document}

\title{A Survey of UV Bright Sources Behind the Halo of M31}

\author[Fittingoff et al.]{
Andrew Fittingoff$^{1}$, 
J. Xavier Prochaska$^{1,2}$,
Jasonjot S. Kalirai$^{2,3}$, 
Jay Strader$^{1,4,5}$ \and
Puragra Guhathakurta$^{1,2}$,
Kyle F. Kaplan$^{1}$\\
$^{1}$Department of Astronomy and Astrophysics,
University of California, Santa Cruz, CA 95064.\\ 
$^{2}$University of California Observatories - 
Lick Observatory, University of California, Santa Cruz, CA 95064\\ 
$^{3}$Space Telescope Science Institute, Baltimore, MD, 21218\\
$^{4}$Harvard-Smithsonian Center for Astrophysics, 
Cambridge, MA 02138\\
$^{5}$Hubble Fellow
}

\maketitle

\begin{abstract}
We have performed a wide-area ultraviolet (UV) imaging survey using the 
GALaxy Evolution eXplorer (GALEX) to search for bright, 
point-like UV sources behind M31's extended halo.  
Our survey consisted of 46 pointings covering an effective
area of $\approx 50$\sqdeg, in both the far-UV and near-UV channels.
We combined these data with optical $R$-band observations acquired with
the WIYN Mosaic-1 imager on the 
Kitt Peak National Observatory 0.9m WIYN telescope.
An analysis of the brightness and colors of sources 
matched between our photometric catalogs
yielded $\approx 100$ UV-bright quasar candidates.
We have obtained discovery spectra for 76 of these targets with
the Kast spectrometer on the Lick 3m telescope and confirm 30
active galactic nuclei and quasars, 29 galaxies 
at $z > 0.02$ including several early-type systems, 
16 Galactic stars (hot main-sequence stars), 
and one featureless source previously identified
as a BL Lac object.  
Future UV spectroscopy of the brightest targets with the 
Cosmic Origins Spectrograph on the Hubble Space Telescope will enable a 
systematic search for diffuse gas in the extended halo of M31.
\end{abstract} 


\section{Introduction}

Structure formation, in current cosmology, follows from the gravitational
collapse of dark matter into virialized `halos'.
Modern theory also predicts that diffuse baryons 
are pulled along during the collapse.  
A subset of these may
travel to the center of the dark matter halo, dissipate their
internal energy, and condense further to form stars. 
A significant fraction, however, may instead
be shock-heated to the virial temperature of the dark matter halo
to form a hot, diffuse baryonic halo.  Modern cosmological simulations
suggest that the amount of baryons that travel along these two paths
is a sensitive function of the total mass, dark matter halo mass, and also
the age of the universe \citep{kkw+05,dbe+08}.
Independent of gravitational heating, feedback processes related to 
galaxy formation (e.g.\ galactic winds, AGN outflows), galaxy-galaxy
interactions (e.g.\ mergers), and even interactions between galaxies and
this diffuse medium (e.g.\ ram-pressure stripping) 
may also pollute the halo with diffuse and, possibly, metal-enriched gas.
Studies of halo gas, therefore, trace the processes of structure
formation and may constrain the roles of feedback in galaxy evolution.

Halo gas is observed in emission from the largest virialized 
structures of our universe: clusters and groups.
Termed the intracluster medium (ICM) or intragroup gas,
this material emits X-rays via Brehmstrahlung processes that are
recorded by space-borne satellites \citep{fj82,mdm+03}.
The observations assess the temperature, enrichment, and density
profiles of this halo gas.  
In turn, the halo's gas serves as a vital diagnostic for the potential
well of massive dark matter halos.  
The analysis indicates
that the majority of baryons in clusters and massive groups are located
within this medium, far surpassing the stellar mass of individual
galaxies \citep[e.g.][]{ars+08}.
Indeed, it has been suggested that diffuse halo gas may be
the dominant baryonic reservoir in the present-day universe
\citep{mulchaey96,fhp98,pt09}.  

%The study of halo gas in emission from lower mass dark matter
%halos (i.e.\ those that host only one or a few galaxies),
%presents a formidable observational challenge.  At
%the virial temperatures characteristic of lower mass systems
%($T_{vir} \lesssim 10^6$\,K), the emission peaks at energies
%that are overwhelmed by a `hot bubble' of foreground, 
%Galactic X-ray emission \citep[e.g.][]{savage79}.
%There are very few reported detections of extended emission
%from halo gas, and even this light is 
%dominated by gas in the inner few tens kpc \citep[e.g.][]{osp07,li08}.  
%Empirical analysis of diffuse halo gas that surrounds isolated
%galaxies, therefore, requires an alternative method.

The most successful approach thus far to studying the halo gas
has been through absorption-line studies,
i.e.\ rest-frame ultraviolet spectroscopy of distant quasars whose
sightlines coincidentally intersect the projected halos of
foreground galaxies.
These observations reveal that galactic halos have a multi-phase medium
comprised of cool, photoionized material 
\citep[traced by $Mg_{II}$ absorption;][]{lb92,kcs+07}
and a warmer, more diffuse component traced by 
higher ionization states \citep[e.g. $C_{IV}$, $O_{VI}$;][]{clw01,tripp08}.
Photoionization modeling of this gas constrains the density, temperature,
and metallicity of the gas \citep[e.g.][]{churchill00a,cp00}.
Because of the inefficiencies of UV spectroscopy, the sample
sizes have been very modest and 
only recently have observers been pursuing studies that yield a statistical
description for a large population of galaxies by focusing
on $Mg_{II}$ transitions redshifted into optical passbands
\citep{zmn+07,ct08}.
Even with this approach, it is extremely rare to study a single 
halo with more than one sightline.  As such, we lack a 
detailed knowledge of the distribution and properties of gas 
(neutral or ionized) within individual galactic halos.

%The one exception is our own Milky Way where
%observations reveal a multi-phase, gaseous halo.  One
%observes a dense, predominantly neutral gas in the form of
%high velocity clouds \citep[e.g.][; HVCs]{pds+02}
%together with a highly ionized, diffuse
%gas traced by \ion{O}{6} and \ion{O}{7} absorption 
%\citep[e.g.][]{sws+03,wangetal05}.
%These emission and absorption-line surveys are limited, however,
%by the poorly constrained distances to the gas owing to our 
%viewing perspective. 
%Ideally, one would perform this
%experiment on an external galaxy, yet this demands a system with 
%large angular size because of the low areal density of 
%UV bright sources on the sky.  

The Andromeda spiral galaxy (M31), a nearby massive spiral, 
is an excellent galaxy for such a study.
Its stellar halo is known to extend out to $\approx 25$ degrees, 
or $\approx 150$kpc \citep{guh05,kgg+06,ggk+06}.  
M31 is a representative late-type spiral of the present Universe.
Furthermore, as our nearest (big galaxy) neighbor there is
a large and expanding dataset of supplementary observations
including multi-band imaging, high spatial resolution images
of select fields, and a tremendous set of optical
spectroscopy on individual stars \citep[e.g.][]{imi+07,grr+06,bsf+07,gfk+07}..
Regarding the gaseous halo, 
clouds of hydrogen gas have been sighted in M31 with similar 
properties as the Galactic HVCs \citep{thilker04}. 
These clouds or their lower column density counterparts are prime targets 
for spectroscopic analysis throughout the extended halo.  
Unfortunately, emission studies of 
warm/hot diffuse gas are challenged by the same foregrounds that
preclude studies within the Milky Way. 
It is a natural extension of our experience with the Galactic halo
to consider surveying M31's halo through absorption-line studies.
The challenge, however, is to identify bright UV sources behind
the halo of M31 and then obtain high fidelity spectroscopy.  

With this challenge as our motivation, we initiated a UV/optical
survey of $\approx 50$\sqdeg\ of M31's galactic halo.
In order to maximize our search area for QSO detections, 
we used the GALaxy Evolution eXplorer (GALEX) telescope \citep{morrissey05}. 
Starting in Cycle~1, the
GALEX team began acquiring images of the disk and inner spheroid of M31.
These data are complemented by images and spectra from
the Sloan Digital Sky Survey.  We were granted a Cycle~2 GO project
to image the extended halo of M31 with GALEX.
This paper reports on these observations and additional optical imaging
and spectroscopy for quasar candidate selection and confirmation.

This paper is organized as follows.  In $\S$~\ref{sec:GALimg},
we describe our GALEX Imaging observations of 
M31's halo.  The optical photometry is discussed in $\S$~\ref{sec:KPNO}
and $\S$~\ref{sec:QSO} details our QSO candidate selection process. 
In $\S$~\ref{sec:spectra} we present first results from our 
follow-up, optical spectroscopy and $\S$~\ref{sec:summ} summarizes
the main findings.

\section{GALEX Imaging}
\label{sec:GALimg}

\begin{figure}
\begin{center}
\includegraphics[scale=0.4]{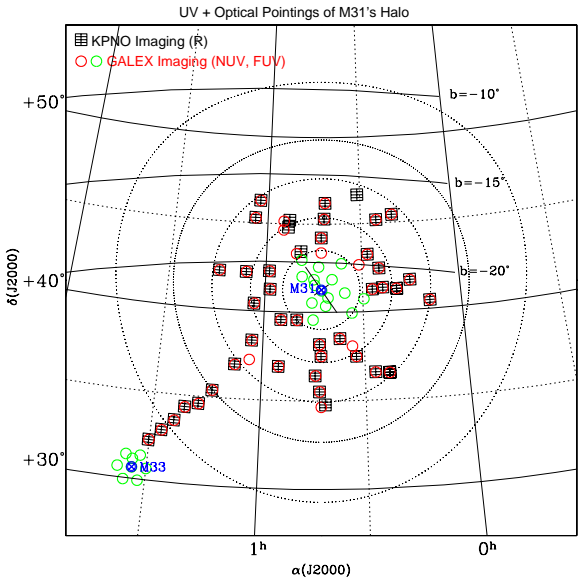}
\end{center}
\caption{Diagram of our Cycle~2 GALEX and our KPNO MOSA
pointings with respect to M31 (GALEX are circles, KPNO are squares).  
Pointings were shifted around to avoid exceptionally bright objects
that could damage GALEX's sensitive instruments.  
The GTO GALEX observations include several pointings on the disk 
of M31 that have also been imaged by the Sloan Digital Sky Survey.
These fields were not included in our program which targets M31's halo.
}
\label{datapts}
\end{figure}

We were awarded a total of 46 pointings during Cycle~2
(PI: Prochaska; ID=033) of the GALEX mission to image a
portion of M31's outer halo.  The pointings 
extend radially outwards from the 
guaranteed time observations
of M31's disk and inner spheroid.  
We had proposed to image the halo along
radial spokes extending to $\sim$\,100~kpc along 
N, NE, E, SE, S, SW, W, and NW directions.  Most of the 
proposed positions were shifted by several
arcminutes to satisfy the GALEX bright star detector limits.
Furthermore, only a subset of our requested pointings were
actually acquired.
These are diagrammed in 
Figure~\ref{datapts} and listed in Table~\ref{galexptngs}.

\begin{table}
 \begin{minipage}{8 truecm}
   \caption{Summary of GALEX pointings. \label{galexptngs} }
  \begin{tabular}{lccc}
  \hline
 ID  & Date (UT) &  $\alpha_{J2000}$  &  $\delta_{J2000}$  \\
  \hline
SW4  & 06 Nov 2005 & 00:32:24.00 & +37:30:00.0 \\
SW5  & 06 Nov 2005 & 00:26:57.18 & +36:35:59.9 \\
S6   & 06 Nov 2005 & 00:42:44.32 & +34:39:00.0 \\
SW3  & 06 Nov 2005 & 00:33:30.78 & +38:05:59.9 \\
S5   & 06 Nov 2005 & 00:43:08.32 & +35:31:01.1 \\
S4   & 06 Nov 2005 & 00:44:24.00 & +36:25:01.1 \\
S3   & 06 Nov 2005 & 00:42:44.32 & +37:31:01.1 \\
SE9  & 07 Nov 2005 & 01:13:47.67 & +35:24:17.9 \\
SW2  & 07 Nov 2005 & 00:37:12.00 & +38:31:47.9 \\
W4   & 08 Nov 2005 & 00:23:49.22 & +41:20:59.9 \\
SE7  & 08 Nov 2005 & 01:07:50.55 & +36:56:41.9 \\
W6   & 08 Nov 2005 & 00:15:18.00 & +41:42:54.0 \\
SE11 & 08 Nov 2005 & 01:21:07.20 & +34:20:59.9 \\
W2   & 08 Nov 2005 & 00:24:42.00 & +42:27:00.0 \\
W3   & 08 Nov 2005 & 00:27:00.00 & +41:16:08.4 \\
E3   & 08 Nov 2005 & 00:58:24.00 & +41:16:08.4 \\
SE2  & 08 Nov 2005 & 00:54:47.99 & +39:33:53.9 \\
NE2  & 09 Nov 2005 & 00:50:33.60 & +43:19:48.0 \\
N1   & 09 Nov 2005 & 00:42:44.32 & +43:22:48.0 \\
E2   & 09 Nov 2005 & 00:58:47.99 & +42:17:59.9 \\
E6   & 09 Nov 2005 & 01:14:24.00 & +42:12:00.0 \\
NE3  & 09 Nov 2005 & 00:54:47.99 & +44:39:00.0 \\
E5   & 09 Nov 2005 & 01:06:00.00 & +42:12:00.0 \\
N3   & 09 Nov 2005 & 00:41:47.99 & +45:17:59.9 \\
N2   & 09 Nov 2005 & 00:42:44.32 & +44:13:47.9 \\
SE12 & 09 Nov 2005 & 01:23:43.35 & +33:32:41.9 \\
NW6  & 10 Nov 2005 & 00:19:49.90 & +45:28:08.4 \\
SE14 & 10 Nov 2005 & 01:30:04.47 & +32:18:18.0 \\
NE6  & 10 Nov 2005 & 01:02:35.99 & +46:17:59.9 \\
SE13 & 10 Nov 2005 & 01:26:53.91 & +32:55:29.9 \\
N4   & 10 Nov 2005 & 00:41:31.20 & +46:12:00.0 \\
NW5  & 10 Nov 2005 & 00:25:00.00 & +45:12:00.0 \\
NE4  & 10 Nov 2005 & 00:54:47.99 & +45:09:00.0 \\
NE5  & 10 Nov 2005 & 01:04:00.00 & +45:17:59.9 \\
NW2  & 29 Jul 2006 & 00:30:56.23 & +42:40:33.8 \\
NW3  & 30 Jul 2006 & 00:28:12.00 & +43:15:00.0 \\
W5   & 30 Jul 2006 & 00:19:33.80 & +41:15:21.8 \\
SW6  & 30 Jul 2006 & 00:22:52.38 & +36:32:49.2 \\
W7   & 06 Aug 2006 & 00:09:36.00 & +40:30:00.0 \\
SE1  & 14 Sep 2006 & 00:50:00.00 & +39:33:53.9 \\
S2   & 14 Sep 2006 & 00:43:12.00 & +38:11:23.9 \\
SE3  & 15 Sep 2006 & 00:55:08.31 & +36:55:29.9 \\
SE6  & 15 Sep 2006 & 01:03:45.43 & +37:15:07.4 \\
E4   & 15 Sep 2006 & 01:03:08.76 & +40:26:13.4 \\
SE5  & 19 Sep 2006 & 01:03:12.00 & +38:20:59.9 \\
SE10 & 19 Sep 2006 & 01:17:22.23 & +34:35:59.9 \\
  \hline
\end{tabular}

Note: The ID represents the direction with respect to the center of M31.
Each pointing was imaged in NUV and FUV, 
with an FOV of $\approx 1.2 {\rm deg}^2$.  
\end{minipage}
\end{table}

\begin{table*}
% \begin{minipage}{8 truecm}
\caption{PHOTOMETRY SUMMARY OF GALEX/$NUV$ DETECTED SOURCES IN THE HALO OF M31\label{tbl_allext_sub}}
\begin{tabular}{rcrrrcccccccccc}
\hline
ID & Name &
$\alpha_{GALEX}$ & $\delta_{GALEX}$ & 
$FUV$ & $\sigma_{FUV}$ & 
$NUV$ & $\sigma_{NUV}$ & 
$R^a$ & $\sigma_{R}$ & 
Class$^b$ & Cand$^c$\\
(J2000) & (J2000) & (AB) & (AB) & (AB) & (AB) & (AB) & (AB)
\\
\hline
     1 & GALEX J000639.12+402641.2 & 00:06:39.12 & +40:26:41.2 & 21.50 &  0.12 & 20.79 & 0.05 & \nodata & \nodata & 0.99& N \\
     2 & GALEX J000639.68+402949.3 & 00:06:39.68 & +40:29:49.3 & 99.99 & -1.00 & 19.54 & 0.02 & \nodata & \nodata & 0.98& N \\
     3 & GALEX J000640.06+403418.4 & 00:06:40.06 & +40:34:18.4 & 99.99 & -1.00 & 19.41 & 0.02 & \nodata & \nodata & 1.00& N \\
     4 & GALEX J000642.73+402405.4 & 00:06:42.73 & +40:24:05.4 & 99.99 & -1.00 & 22.16 & 0.10 & \nodata & \nodata & 1.00& N \\
     5 & GALEX J000643.10+402835.3 & 00:06:43.10 & +40:28:35.3 & 99.99 & -1.00 & 20.78 & 0.04 & \nodata & \nodata & 0.98& N \\
     6 & GALEX J000644.53+403139.9 & 00:06:44.53 & +40:31:39.9 & 21.11 &  0.10 & 20.60 & 0.04 & \nodata & \nodata & 1.00& N \\
\hline
\end{tabular}
Note: 
a: $R$-band magnitudes from KPNO photometry are reported for the NUV extraction that are matched to within a $2''$ radius of a KPNO source.  For NUV extractions where KPNO imaging exists but no source was detected, we report a value of 99.99 and an error of -1.  For NUV extractions where no KPNO imaging exists, we report no values.
a: Star/galaxy classifier calculated by SExtractor on the NUV photometry.  Values near unity indicate a stellar-like point-spread function. 
b: Quasar candidates picked based on color relations between UV and R.  The criteria used were $.5<FUV-NUV<2$, $NUV-R<$2, and $NUV<21$. 
d: NUV extractions with no FUV match have FUV values of 99.99 and an error of -1.
[The complete version of this table is in the electronic edition.  The printed edition contains only a sample.]
%\end{minipage}
\end{table*}

The exposure time for each observation was set to one
orbit (1631~s). The incoming beam is split by
a dichroic into the far-UV ($FUV; 1350-1790$\AA) and 
near-UV ($NUV; 1770-2730$\AA) channels.  The unvignetted field-of-view
of each pointing is a circle with radius $\approx 35'$ that is imaged
simultaneously by two cameras for a total field-of-view (FOV)
of approximately 1.2\,\sqdeg.
The raw images were processed through the reduction pipeline at
the GALEX Science Operations 
Center\footnote{http://galexgi.gsfc.nasa.gov/docs/galex/Documents/GALEXPipelineDataGuide.pdf}.
The output image has been background subtracted, flux calibrated, and
astrometrically corrected to an accuracy of 1$''$.

We measured the brightness of all sources on each image, in both
filters, using the SExtractor software package
\citep[version 2.5.0;][]{bertin96}.  The full width at half maximum
(FWHM) of point sources on each \nuv\ image subtends $\sim$3.2
pixels.  Given the pixel scale of the detector (1.5$''$ pixel$^{-1}$),
the angular FWHM is 4.8$''$ (the FWHM for the \fuv\ exposures is slightly
smaller at 4.2$''$).  The aperture size was set to the FWHM, and each source
with counts more than $\sim 3 \sigma$ above the mean 
local sky-subtracted background was
considered a detection.  The resulting catalog included a large number of
artifacts at the circular edges of each image.  We trimmed the catalogs
to only include a fraction of the total radius, i.e.\  0.5625 deg
from the center (1325 pixels) giving a total area 0.994 deg$^2$ for
each field.  This resulted in a total number of \numnuv\ objects in the
combined \nuv\ data.  Approximately $\sim$15\% of the sources were also detected
at $3\sigma$ significance in the \fuv\ images.
The sources that were undetected in the \fuv\ images were retained in the
final catalogs 
but flagged to note that a precise $FUV-NUV$
color is not available.
The sources without significant \fuv\ flux are 
unlikely to correspond to UV-bright sources background to M31.
The fixed-aperture photometry for all sources detected in the \nuv\ band is presented
in Table~\ref{tbl_allext_sub}.

\begin{figure}
\begin{center}
\includegraphics[scale=0.3,angle=90]{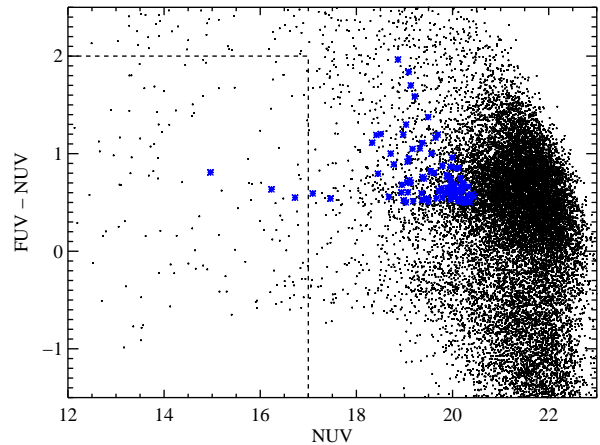}
\end{center}
\caption{Color-magnitude diagram of all GALEX objects.  
Our initial criteria for picking out QSOs was to look for 
bright \nuv\ magnitudes and blue $FUV-NUV$ color (within the box on the diagram).
Follow-up spectroscopy indicates that a very high fraction of these
sources are foreground Galactic stars.  As such, we obtained
KPNO optical imaging of the fields to perform more efficient
color-color pre-selection.  The blue stars indicate the position of
these preferred candidates in the NUV/FUV color-magnitude diagram.
}
\label{cmd2}
\end{figure}

The reduced images from the GALEX pipeline record the data in flux units
where one digital number (DN) corresponds to
$1.4 \times 10^{-15} {\rm erg \; s^{-1} \; cm^{-2} \; \AA^{-1}}$
for the \fuv\ images and
$2.06 \times 10^{-16} {\rm erg \; s^{-1} \; cm^{-2} \; \AA^{-1}}$
for the \nuv\ images.
The conversion from flux to AB magnitudes for GALEX data is given 
by\footnote{http://www.galex.caltech.edu/researcher/techdoc-ch2.html}
\begin{equation}
M_{FUV} = 18.82 - 2.5 * \log(DN)
\end{equation}
\begin{equation}
M_{NUV} = 20.08 - 2.5 * \log(DN)
\end{equation}
where 18.82 and 20.08 are the corresponding zero point AB magnitudes.
A color magnitude diagram of all $\approx 17,000$ sources detected in both filters
in our survey is presented in Figure~\ref{cmd2}.

%We identified 137 objects that
%satisfied our color criterion and had $NUV < 17$\,mag.  
%We then obtained follow-up spectroscopy with the Lick/KAST spectrometer 
%(see $\S$~\ref{sec:spectra}) in Fall 2006
%for a subset of these sources and were disappointed to find 
%that all of them were stellar in nature (B type main-sequence stars).  
%We concluded that the false-positive
%detection for candidates based on UV photometry alone was too high for
%an effective follow-up program.  
%Therefore, we proceeded to obtain optical
%photometry overlapping our GALEX images in M31's halo to lengthen
%our color baseline.

%\input{../tbl_kpnoptns_mnras.tex}
\begin{table}
 \begin{minipage}{8 truecm}
\caption{Log of KPNO Observations. \label{kpnoobs} }
\begin{tabular}{ccccc}
  \hline
 ID  & Date & $\alpha_{J2000}$ & $\delta_{J2000}$ & FWHM \\ 
& (UT) &&&(pix) \\ 
  \hline
N2   & 25 Nov 2006 & 00:42:43.82 & 44:13:51.18 & 2.1 \\
N3   & 25 Nov 2006 & 00:41:47.79 & 45:18:04.08 & 2.2 \\
N4   & 25 Nov 2006 & 00:41:30.95 & 46:12:03.05 & 2.1 \\
NW2  & 25 Nov 2006 & 00:30:56.26 & 46:40:35.03 & 2.1 \\
NW3  & 25 Nov 2006 & 00:28:12.22 & 43:15:04.82 & 2.1 \\
NW5  & 25 Nov 2006 & 00:25:00.27 & 45:12:02.60 & 2.2 \\
NW6  & 25 Nov 2006 & 00:19:50.17 & 45:28:11.29 & 2.1 \\
W2   & 25 Nov 2006 & 00:24:42.47 & 42:27:05.58 & 2.2 \\
W3   & 25 Nov 2006 & 00:27:00.63 & 41:16:16.73 & 2.5 \\
W4   & 25 Nov 2006 & 00:23:49.94 & 41:21:06.52 & 2.4 \\
W5   & 25 Nov 2006 & 00:19:34.69 & 41:15:28.28 & 2.3 \\
W5   & 25 Nov 2006 & 00:19:32.88 & 41:15:29.22 & 2.0 \\
W6   & 25 Nov 2006 & 00:15:18.02 & 41:43:00.03 & 2.1 \\
W7   & 25 Nov 2006 & 00:09:36.34 & 40:30:06.84 & 2.0 \\
SW2  & 25 Nov 2006 & 00:37:12.29 & 38:31:56.55 & 1.7 \\
SW4  & 25 Nov 2006 & 00:32:24.28 & 37:30:10.10 & 1.7 \\
SW5  & 25 Nov 2006 & 00:26:57.67 & 36:36:10.68 & 1.4 \\
SW6  & 25 Nov 2006 & 00:22:52.89 & 36:32:59.41 & 1.5 \\
SW6  & 25 Nov 2006 & 00:22:56.29 & 36:33:06.41 & 1.9 \\
SW2  & 25 Nov 2006 & 00:22:56.29 & 36:33:06.41 & 1.8 \\
S2   & 25 Nov 2006 & 00:43:17.61 & 38:11:48.18 & 2.0 \\
S3   & 25 Nov 2006 & 00:42:49.96 & 37:31:27.13 & 2.0 \\
S4   & 25 Nov 2006 & 00:44:29.66 & 36:25:28.32 & 1.8 \\
S5   & 25 Nov 2006 & 00:43:14.14 & 35:31:31.18 & 1.8 \\
SE1  & 25 Nov 2006 & 00:50:06.02 & 39:34:20.02 & 1.7 \\
SE2  & 25 Nov 2006 & 00:54:54.16 & 39:34:19.68 & 1.6 \\
SE3  & 25 Nov 2006 & 00:55:14.42 & 36:55:59.73 & 1.9 \\
E2   & 25 Nov 2006 & 00:58:54.44 & 42:18:25.27 & 1.9 \\
SE5  & 25 Nov 2006 & 01:03:18.28 & 38:21:27.97 & 1.6 \\
SE7  & 25 Nov 2006 & 01:07:56.74 & 36:57:11.77 & 2.0 \\
SE9  & 25 Nov 2006 & 01:13:53.69 & 35:24:48.86 & 1.8 \\
E6   & 25 Nov 2006 & 01:14:30.60 & 42:12:24.98 & 1.7 \\
SE11 & 25 Nov 2006 & 01:21:13.28 & 34:21:31.33 & 2.0 \\
SE12 & 25 Nov 2006 & 01:23:49.48 & 33:33:13.89 & 1.7 \\
SE13 & 25 Nov 2006 & 01:27:00.06 & 32:56:03.59 & 1.8 \\
SE14 & 25 Nov 2006 & 01:30:10.64 & 32:18:52.31 & 1.8 \\
S6   & 28 Nov 2006 & 00:41:33.84 & 34:47:32.00 & 1.7 \\
NE2  & 28 Nov 2006 & 00:49:09.75 & 43:26:47.30 & 1.5 \\
NE3  & 28 Nov 2006 & 00:53:17.98 & 44:49:43.30 & 1.8 \\
NE4  & 28 Nov 2006 & 00:52:58.79 & 45:13:39.10 & 1.4 \\
E3   & 28 Nov 2006 & 00:58:29.82 & 41:16:30.15 & 2.0 \\
E4   & 28 Nov 2006 & 01:03:14.52 & 40:26:35.91 & 1.9 \\
E5   & 28 Nov 2006 & 01:06:05.88 & 42:12:21.39 & 1.6 \\
NE6  & 28 Nov 2006 & 01:02:42.45 & 46:18:18.82 & 2.0 \\
NE5  & 28 Nov 2006 & 01:04:06.49 & 45:18:20.05 & 1.9 \\
SE10 & 28 Nov 2006 & 01:17:27.75 & 34:36:28.51 & 2.0 \\
  \hline
\end{tabular}
Note: The ID represents the direction with respect to the center of M31.
Observations were taken with an R-band filter (6200-7500$\AA$) with an 
$\approx .93 {\rm deg}^2$ FOV and an exposure time of 60s.
The pointings are intended to overlap with the data we received from GALEX.  The FWHM values
were estimated from the images directly.
\end{minipage}
\end{table}

\section{KPNO Imaging}
\label{sec:KPNO}

%To identify quasar candidates, we examined
%the $FUV - NUV$ color-magnitude diagram shown below for bright ($NUV < 17$)
%and blue ($FUV-NUV<2$) objects. This range of
%magnitudes and colors has been successfully used to find quasars through
%the matching of SDSS and GALEX observations \citep{bianchi05}.
%The overwhelming majority of the Galactic stellar locus, our principle
%contaminant, lies at
%significantly fainter \nuv\ magnitudes and redder $FUV - NUV$ colors.
%For a perfect blackbody spectrum, an $FUV-NUV = 2$ color corresponds to
%an effective temperature $T$ = 8000 K.  

To increase the efficiency of selecting quasar
candidates, we chose to extend our baseline to include
optical photometry in the $R$-band \citep[e.g.][]{atlee07}.
Although our GALEX data revealed many thousands of bright and blue UV sources,
these data by themselves do not provide an efficient means of selecting
quasar candidates.   
%One can significantly improve the selection
%efficiency by obtaining optical photometry \citep[e.g.][]{atlee07}.  
We acquired $R$-band images using the WIYN Mosaic-1 imager on the
KPNO 0.9 meter WIYN telescope.  This section describes the observations,
data reduction, and analysis of these images.

\subsection{Image Reduction}

On the nights of 25--28 November 2006, we acquired 43 $R$-band images covering
nearly the entire area of our GALEX imaging program.
We used the WIYN Mosaic-1 imager which records the data in a mosaic of
eight CCDs, each one with a viewing area of $14.49' \times 28.97'$
($0.43''$\,pixel$^{-1}$).  Therefore, each mosaic image has a total area of
$\approx .93 {\rm deg}^2$, similar to each of our GALEX images.
A single Mosaic-1 frame overlapping each GALEX image was obtained with an exposure 
time of 60 seconds.  The nights were relatively clear but
thin cirrus precludes precise photometric calibration
from standard star fields observed during this run. 
The seeing ranged from 1 to 2$''$,
and the airmass was $<$1.6 for all observations.

The $R$ band images were reduced using standard IRAF tasks in the
\textbf{mscred} module.  We first calculated coefficients 
for each chip of the mosaic data to account for cross-talk between 
adjacent CCDs.  We applied the bad pixel masks, and then subtracted a 
mean bias frame from each of the raw images.  Finally, we corrected for 
pixel-to-pixel variations by dividing the images by a combined dome flat field.  
The resulting images, over the full mosaic, showed significant sky gradients
caused by spatial and temporal variations in our sky flats.  To remove
the gradients, we used the IRAF task \textbf{mscskysub} to fit a 
surface to each CCD and subtract the mean background value.  In this process, 
the subtraction ignored large residuals which represent actual objects on
the images.  The \textbf{mscskysub} task is typically used to smooth out
the background sky after stitching together individual CCDs to create
a large mosaic image.  Although we executed this task on individual
chips, we checked and confirmed that the photometry derived from the
images with and without this processing showed very little difference
($<10\%$) in the results, provided one performs local sky subtraction.
Cosmic rays were not removed or flagged, but the images were short
exposures and therefore would have a small contamination rate.

\subsection{Astrometry and Photometry}

We measured the positions and magnitudes of all sources separately
on each of the eight CCDs of the mosaic.  Because our observational
design consists of a single exposure of each field, this leads to 368 total
sub-images.  The world coordinate system information recorded in the image
headers was systematically in error.  We therefore submitted each of the 368 
images to Astrometry.net \citep{astrometrynet}
to yield a better estimate of the field center and overall
solution.  The resulting solution provided an adequate fit over most of
the field of view, but some residual mismatch was still seen towards the
corners of the chips.  The solution was iterated using the \textbf{wcstools}
module within IRAF by matching only a subset of the brightest, well measured
stars to the USNO-B1.0 catalog.  The fitting function was a quadratic
polynomial after the initial offsets, scales, and rotations were calculated.
Recentering of the image center was allowed, as well as sigma clipping of
objects that were more than 2$\sigma$ from the expected location based on
the previous fit.  For the best frames, the rms residual of the solution was
less than 0.3$''$ based on several dozen guide stars.  
For a few of the frames,
less than 10 stars were found in common between our catalog and the USNO
and we could not recover a useful astrometric soluation;  these
frames were ignored in all subsequent analysis.

Point-spread function (PSF)
photometry was performed on each processed image using DAOPHOT and
ALLSTAR \citep{stetson94}.  As the seeing varied from 1$''$ to 2$''$ during
the observing run, the fitting radius was set separately for each image
to be 80\% of the image FWHM.  This was measured using the \textbf{imexam}
IRAF task for a number of bright, isolated, non-saturated stars (see
Table~\ref{kpnoobs}).  All sources more than 3$\sigma$ above the local
background were considered positive detections.  A PSF template was calculated
using $\sim$50 bright stars on each frame, and fit to and subtracted from each 
source to create a residual image.  A second pass of the photometry was made on
this residual image to yield the final catalog.

We calibrated the magnitudes by comparing to the 
USNO catalog, which has an $R$ magnitude and
a $B-R$ color.  The overlap between our optical source catalog and the USNO
catalog ranged from $R$ = 10 to 19.  We fit for a color term and offset
for each CCD individually, after applying a cut to eliminate outliers that
were more than 2$\sigma$ from the mean.   We note
that the offsets varied by several tenths of a magnitude for different CCDs.
We found that the typical limiting magnitude for our 
$3\sigma$ detection criterion is $R \approx 20$.
The $R$-band photometry of all KPNO sources that were matched 
to a GALEX \nuv\ detection
are listed in Table~\ref{tbl_allext_sub}.

\section{QSO Candidates}
\label{sec:QSO}

\subsection{Catalog Combination}

Our total catalog of GALEX detections via SExtractor,
after eliminating duplicates, 
contained \nodupg\ objects, which were determined by
detection in the NUV.  Duplicate detections were chosen
to be within $0.5''$, and the magnitudes taken from
the entry with the smallest error.  
Objects without a corresponding
FUV detection were given a magnitude value of 99.99 and an
error of -1.
The total KPNO catalog,
after eliminating erroneous detections, totalled
144,531 objects.
%The discrepency is most likely due to stellar sources
%being more likely to be detected in $R$-band than in the UV.  
To find matches between the 
catalogs we tested for the nearest object within $1.5''$. 
The matched GALEX+KPNO catalog
contained \numgk\ objects.  
Given the low source density of objects in our relatively
shallow $R$-band images, we estimate that the
number of spurious matchees is fewer than 200.

\begin{figure}
\begin{center}
\includegraphics[scale=0.3,angle=90]{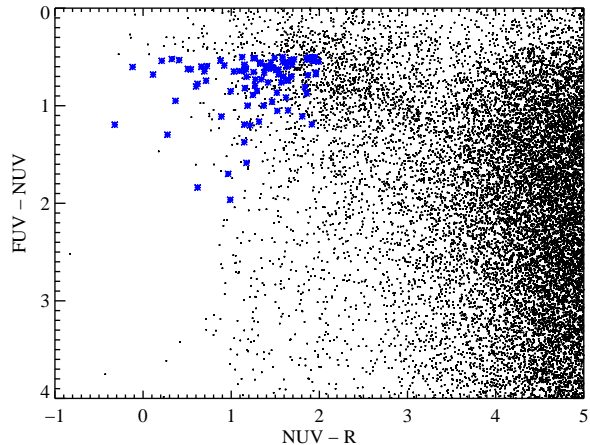}
\end{center}
\caption{Color-color diagram of all extracted objects 
detected in each of our $FUV$, $NUV$, and KPNO/$R$ imaging
(AB magnitudes).  
QSO candidates (blue) were determined by location in color-color space 
and with \nuv\ and $FUV < 21$.
The black dots that lie in the region of color-color space occupied by our preferred 
QSO candidates do not satisfy our brightness criteria $(NUV<21$).
}
\label{fig:clrclr}
\end{figure}

\subsection{Candidates}

The criteria we used to define a quasar candidate list were taken
from previous papers on QSO detections using a combination of
GALEX and optical data \citep{bianchi07,atlee07}.  
The general approach is to take the area in color-color space 
which maximizes the density of QSOs while minimizing contaminants.  
QSO colors that include GALEX observations 
have been measured using the SDSS database.
The selections fall into an area defined 
by searching within these limiting values: 
$0.5 < FUV-NUV < 2$, $NUV-R<2$, and $NUV < 21$.
The process, as done by \cite{bianchi07}, gives an 85\% success rate
in comparing QSOs identifed by color and those spectroscopically
confirmed.  This is consistent with our own results, 
as detailed in $\S$~\ref{sec:spectra}.

This yielded a primary quasar candidate list of 97 targets
(Table \ref{tbl_allext_sub}, Figure~\ref{fig:clrclr}).

\begin{table}
 \begin{minipage}{8 truecm}
\caption{Lick/KAST Spectroscopy of QSO Candidates\label{tab:lick}}
\begin{tabular}{cccccc}
\hline
Name & $\alpha_{GALEX}$ &$\delta_{GALEX}$ &Date &Exp.  \\
& (J2000)& (J2000) & (UT) & (s) 
\\
\hline
J000817+400223&00:08:17&+40:02:23&28Sep2008&600\\
J000825+400345&00:08:25&+40:03:45&28Sep2008&600\\
J001036+400314&00:10:36&+40:03:14&25Sep2008&600\\
J001117+402202&00:11:17&+40:22:02&26Sep2008&800\\
J001502+412756&00:15:02&+41:27:56&28Sep2008&600\\
J001553+412026&00:15:53&+41:20:26&25Sep2008&600\\
J001601+411820&00:16:01&+41:18:20&29Sep2008&720\\
J001632+414752&00:16:32&+41:47:52&26Sep2008&800\\
J001836+411611&00:18:36&+41:16:11&28Sep2008&300\\
J002007+450802&00:20:07&+45:08:02&26Sep2008&900\\
\hline
\end{tabular}
[The complete version of this table is in the electronic edition.  The printed edition contains only a sample.]
\end{minipage}
\end{table}

\begin{figure}
\begin{center}
\includegraphics[scale=0.4]{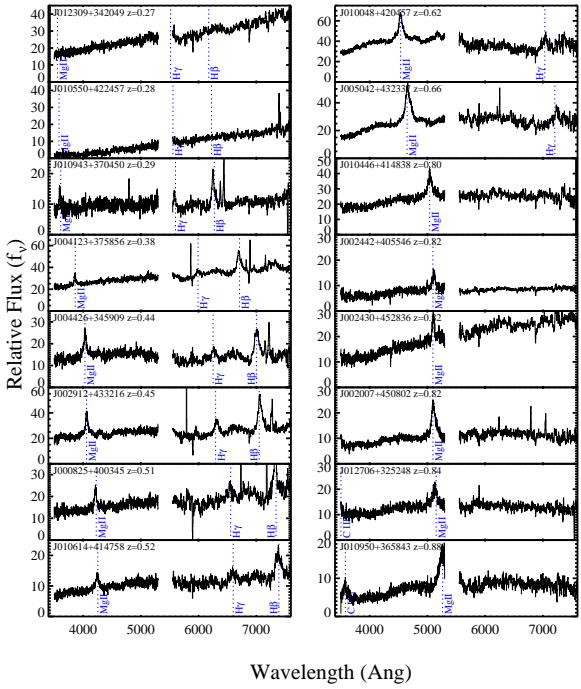}
\end{center}
\caption{Lick/Kast discovery spectra of UV-bright AGN and quasars
in the field surrounding M31.  The data at $\lambda < 5400$\AA\ ($>5600$\AA)
correspond to the blue (red) camera of the Kast spectrometer.
The gap at $\lambda \approx 5400-5500$\AA\ corresponds to the central
wavelength of the dichroic used in the observations.
The dotted vertical lines show the positions of prominent emission
lies for AGN, shifted according to the measured redshift.
}
\label{fig:agn}
\end{figure}

\begin{figure}
\begin{center}
\includegraphics[scale=0.4]{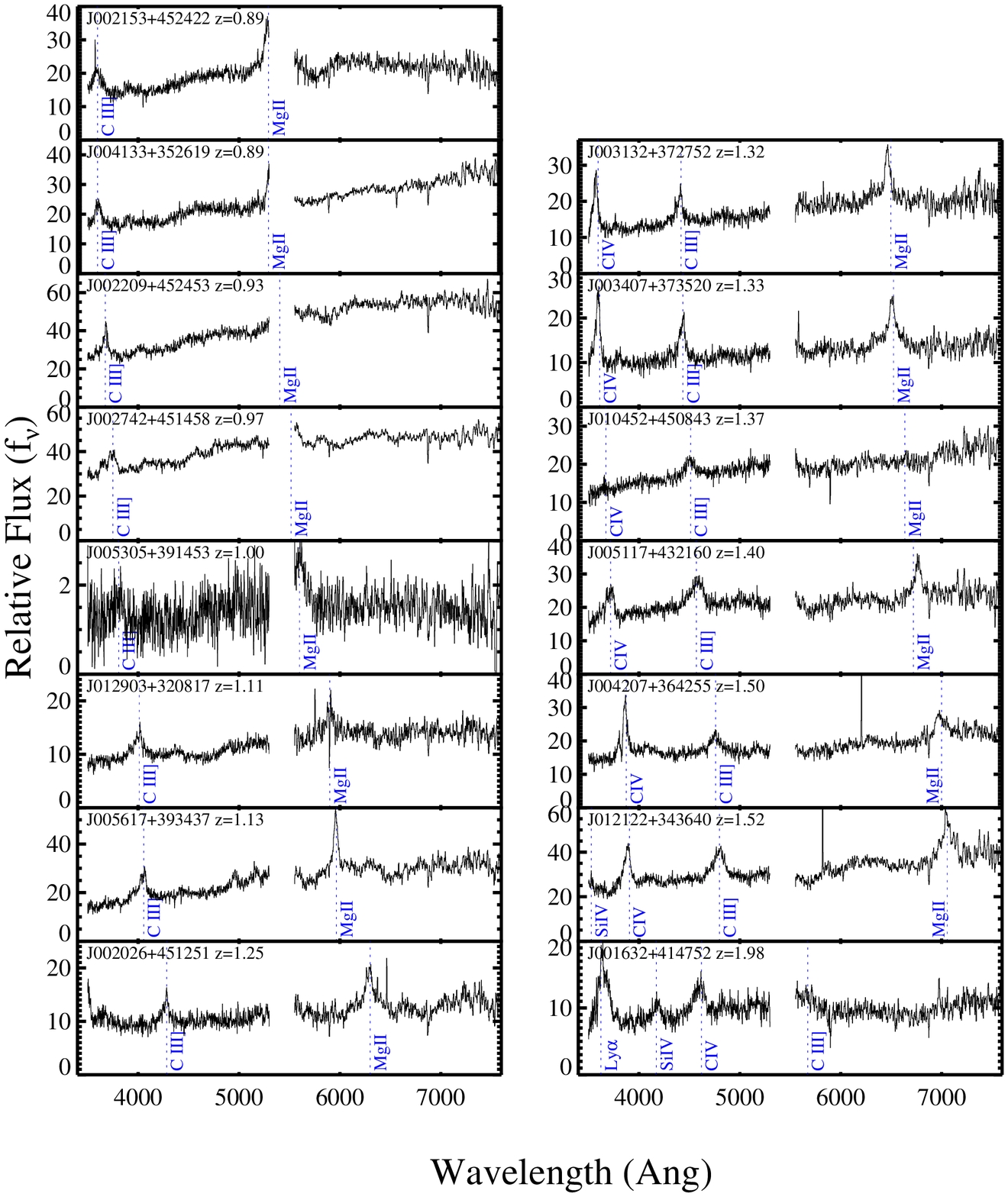}
\end{center}
\end{figure}

\begin{figure}
\begin{center}
\includegraphics[scale=0.4]{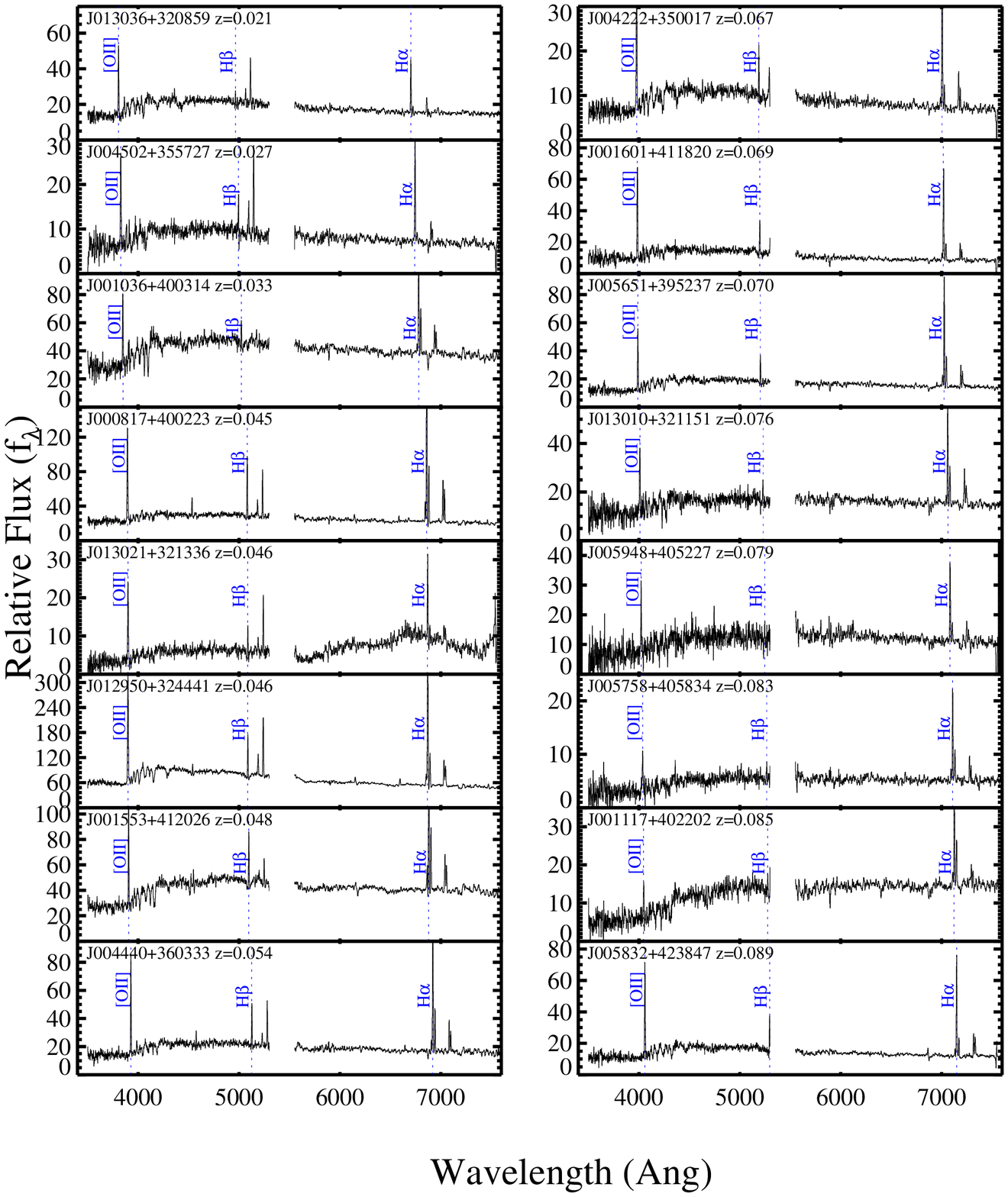}
\end{center}
\caption{Same as Figure~\ref{fig:agn} but for extragalactic objects
showing only narrow emission lines, presumed to be star-forming galaxies.
}
\label{fig:gal}
\end{figure}

\begin{figure}
\begin{center}
\includegraphics[scale=0.4]{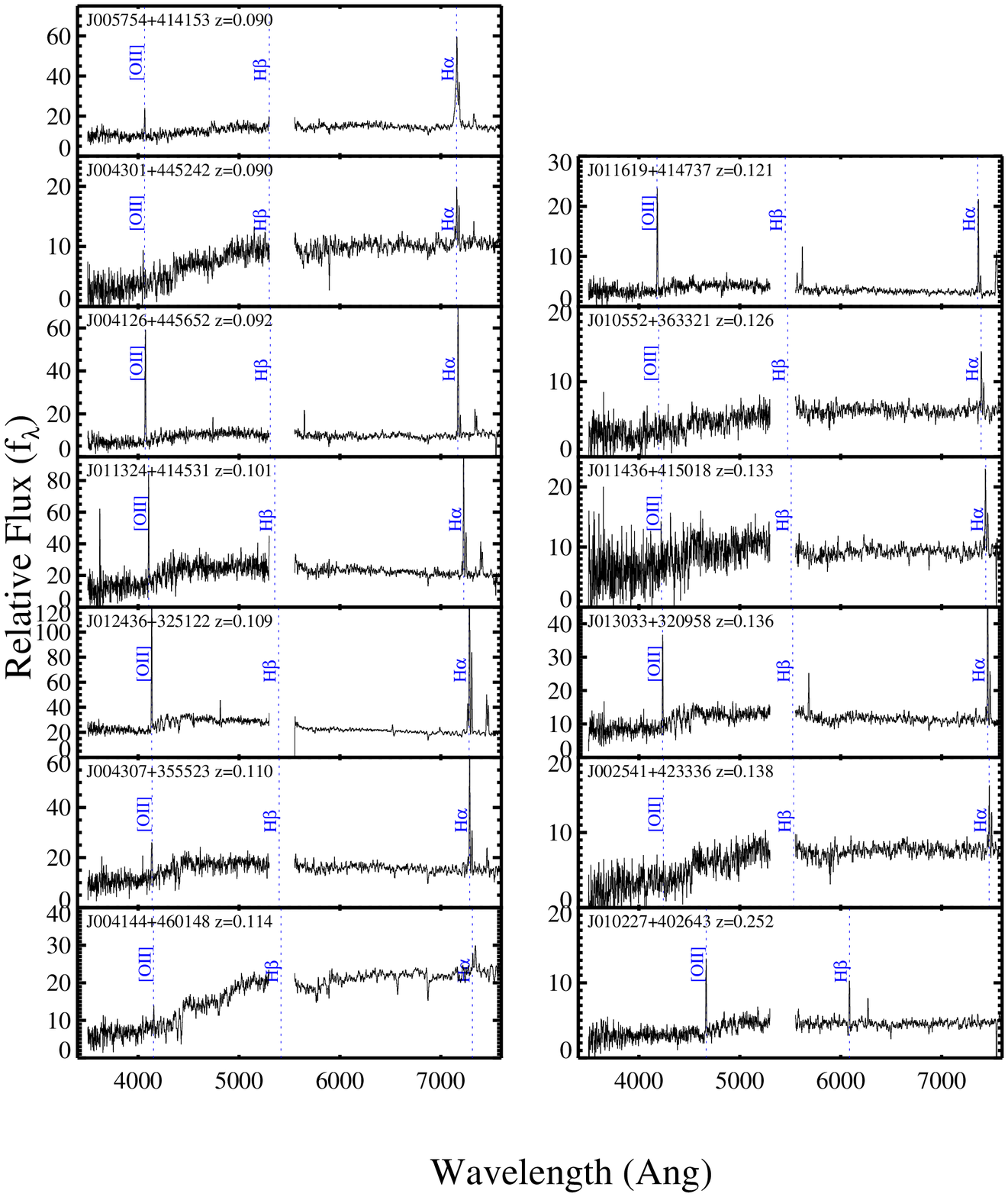}
\end{center}
\end{figure}

\section{Discovery Spectra}
\label{sec:spectra}

We obtained spectra for a handful of our candidates during unrelated 
observing runs in Winter 2008 at the Keck and Lick Observatories
with the LRIS \citep{lris}
and Kast spectrometers respectively.
Several of these spectra revealed extragalactic sources including
two AGN.  We then initiated a spectroscopic follow-up campaign
at Lick Observatory.  The first results of this survey are presented here.

Table~\ref{tab:lick} summarizes the observations of QSO candidates 
obtained during a five
night run at Lick Observatory (September 25$-$29, 2008).
These were drawn primarily from the candidate list given
in Table~\ref{tbl_allext_sub}, but were supplemented by targets with slightly
less stringent color criteria.
For all observations we used
the Kast spectrograph mounted on the Lick 3m telescope.
The instrument was configured with the $2''$-wide slit and
the D55 dichroic that split the light into the blue and red cameras.
We employed the 452/3306 grism on the blue side giving
a FWHM$\approx 2.8$\AA\ spectral resolution for wavelengths spanning
3300$-$5400\AA.  On the red side, we used the 600/7500 grating tilted
to provide coverage from 5600$-$8000\AA\ with a 
FWHM$\approx 4.6$\AA\ spectral resolution.
We took spectral images of Ar, He, Cd, and quartz lamps for 
wavelength calibration and flat-fielding.

\begin{table*}
 \begin{minipage}{8 truecm}
\caption{UV Bright, Extragalactic Sources Behind M31 \label{tab:extra}}
\begin{tabular}{ccccccc}
\hline
Name & $\alpha_{GALEX}$&$\delta_{GALEX}$&$\rho_{\rm M31}^a$&$FUV$&$NUV$&$z$\\ 
& (J2000)& (J2000) & (kpc) & (AB) & (AB) 
\\
\hline
\multicolumn{7}{c}{AGN (Broad Emission Lines)}\\
\hline
J000825+400345&00:08:25&+40:03:45& 90&19.14&18.78& 0.51\\
J001632+414752&00:16:32&+41:47:52& 67&20.53&19.97& 1.98\\
J002007+450802&00:20:07&+45:08:02& 77&20.93&19.09& 0.82\\
J002026+451251&00:20:26&+45:12:51& 77&20.36&19.54& 1.25\\
J002153+452422&00:21:53&+45:24:22& 77&20.83&18.86& 0.89\\
J002209+452453&00:22:09&+45:24:53& 76&20.98&20.32& 0.93\\
J002430+452836&00:24:30&+45:28:36& 73&20.63&19.97& 0.82\\
J002442+405546&00:24:42&+40:55:46& 47&20.80&19.22& 0.82\\
J002742+451458&00:27:42&+45:14:58& 66&19.24&18.45& 0.97\\
J002912+433216&00:29:12&+43:32:16& 46&19.22&18.96& 0.45\\
J003132+372752&00:31:32&+37:27:52& 60&19.83&19.10& 1.32\\
J003407+373520&00:34:07&+37:35:20& 55&20.37&19.32& 1.33\\
J004123+375856&00:41:23&+37:58:56& 45&19.14&18.73& 0.38\\
J004133+352619&00:41:33&+35:26:19& 80&19.77&19.06& 0.89\\
J004207+364255&00:42:07&+36:42:55& 62&20.81&19.65& 1.50\\
J004426+345909&00:44:26&+34:59:09& 86&19.39&19.10& 0.44\\
J005042+432337&00:50:42&+43:23:37& 35&19.71&18.71& 0.66\\
J005117+432160&00:51:17&+43:21:60& 36&20.48&19.37& 1.40\\
J005305+391453&00:53:05&+39:14:53& 39&20.89&19.69& 1.00\\
J005617+393437&00:56:17&+39:34:37& 42&19.98&19.06& 1.13\\
J010048+420457&01:00:48&+42:04:57& 47&19.67&18.78& 0.62\\
J010446+414838&01:04:46&+41:48:38& 57&19.71&18.51& 0.80\\
J010452+450843&01:04:52&+45:08:43& 76&20.14&19.38& 1.37\\
J010550+422457&01:05:50&+42:24:57& 61&19.70&19.18& 0.28\\
J010614+414758&01:06:14&+41:47:58& 60&20.34&19.72& 0.52\\
J010943+370450&01:09:43&+37:04:50& 91&19.86&19.63& 0.29\\
J010950+365843&01:09:50&+36:58:43& 92&20.87&19.49& 0.88\\
J012122+343640&01:21:22&+34:36:40&137&20.21&19.17& 1.52\\
J012309+342049&01:23:09&+34:20:49&143&19.24&18.91& 0.27\\
J012706+325248&01:27:06&+32:52:48&165&19.78&19.35& 0.84\\
J012903+320817&01:29:03&+32:08:17&176&20.56&19.56& 1.11\\
\hline
\multicolumn{7}{c}{Galaxies (Narrow Emission Lines)}\\
\hline
J000817+400223&00:08:17&+40:02:23& 91&19.38&18.94&0.045\\
J001036+400314&00:10:36&+40:03:14& 85&19.23&18.68&0.033\\
J001117+402202&00:11:17&+40:22:02& 82&20.83&19.96&0.085\\
J001553+412026&00:15:53&+41:20:26& 69&19.53&19.03&0.048\\
J001601+411820&00:16:01&+41:18:20& 68&19.68&19.31&0.069\\
J002541+423336&00:25:41&+42:33:36& 47&20.63&20.11&0.138\\
J004126+445652&00:41:26&+44:56:52& 50&20.66&20.03&0.092\\
J004144+460148&00:41:44&+46:01:48& 65&20.67&19.79&0.114\\
J004222+350017&00:42:22&+35:00:17& 85&20.57&20.01&0.067\\
J004301+445242&00:43:01&+44:52:42& 49&20.34&19.79&0.090\\
J004307+355523&00:43:07&+35:55:23& 73&20.20&19.66&0.110\\
J004440+360333&00:44:40&+36:03:33& 71&19.89&19.37&0.054\\
J004502+355727&00:45:02&+35:57:27& 73&19.92&19.46&0.027\\
J005651+395237&00:56:51&+39:52:37& 41&20.66&19.98&0.070\\
J005754+414153&00:57:54&+41:41:53& 39&20.71&20.04&0.090\\
J005758+405834&00:57:58&+40:58:34& 39&20.76&20.26&0.083\\
J005832+423847&00:58:32&+42:38:47& 44&20.65&20.04&0.089\\
J005948+405227&00:59:48&+40:52:27& 44&20.86&20.36&0.079\\
J010227+402643&01:02:27&+40:26:43& 52&20.73&20.19&0.252\\
J010552+363321&01:05:52&+36:33:21& 89&20.69&19.95&0.126\\
J011324+414531&01:13:24&+41:45:31& 79&20.75&20.15&0.101\\
J011436+415018&01:14:36&+41:50:18& 82&20.96&20.38&0.133\\
J011619+414737&01:16:19&+41:47:37& 86&20.91&20.16&0.121\\
J012436+325122&01:24:36&+32:51:22&161&19.83&19.13&0.109\\
J012950+324441&01:29:50&+32:44:41&172&18.54&18.04&0.046\\
J013010+321151&01:30:10&+32:11:51&178&20.79&20.26&0.076\\
J013021+321336&01:30:21&+32:13:36&178&20.97&20.12&0.046\\
J013033+320958&01:30:33&+32:09:58&179&20.37&19.75&0.136\\
J013036+320859&01:30:36&+32:08:59&179&19.54&18.94&0.021\\
\hline
\end{tabular}
Note: a: Impact parameter from the center of M31, taken to be $\alpha_{M31} = $00:42:44.3, $\delta_{M31}$= +41:16:09 (J2000), for an assumed distance of 783\,kpc.
\end{minipage}
\end{table*}

The data were reduced with the 
LowRedux\footnote{http://www.ucolick.org/$\sim$xavier/LowRedux/index.html}
package
developed by J. Hennawi, S. Burles, and JXP that is bundled within
the XIDL\footnote{http://www.ucolick.org/$\sim$xavier/IDL/}
software package.
The data were flat fielded to remove pixel-to-pixel variations,
wavelength calibrated, sky subtracted, and optimally extracted.
The 1D spectra were corrected for instrumental flexure
and then flux calibrated using a sensitivity function derived from observations
of a spectrophotometric standard taken the first night.  All nights were clear and likely
to have been nearly photometric.  The absolute flux calibration of the
data is not accurate because of slit losses but we estimate the relative fluxing
is accurate to within $\approx 20\%$.  For targets with multiple exposures,
we coadded the 1D spectra after weighting by the median inverse variance.

Each 1D spectrum of our quasar candidates
was visually analyzed for significant absorption and
emission lines to perform object classification.
The principle contaminants to the quasar candidates are:
(i) foreground Galactic stars that show prominent Balmer absorption lines and
(ii) star-forming galaxies that exhibit narrow emission lines and
occasional CaH+K and Balmer absorption lines.
In nearly every spectrum, we identified multiple emission and/or absorption
features which unambiguously classified the object and yielded an
estimate of the target's redshift.  All of the extragalactic sources
are listed in Table~\ref{tab:extra}.  We characterize objects with
narrow emission lines as galaxies and report all others as AGN 
(many of these may be more properly classified as quasars).
There is one source (J012309+342049) that exhibits a nearly featureless spectrum
across our full wavelength coverage.   The absence of Balmer absorption
features rules out this target
as a Galactic star.  We initially classified it as 
an unknown extragalactic source and
suspected it to be a blazar.  Indeed, a search using SIMBAD
revealed this source was previously known
as a BL Lacertae object at $z=0.27$ \citep{pss+96}.  
One of the other AGN
was also previously discovered using 
X-ray imaging \citep[J002742+451457;][]{xcb+97}.
Spectra of all the extragalactic sources are shown 
in Figures~\ref{fig:agn} and \ref{fig:gal}.

\begin{figure}
\begin{center}
\includegraphics[scale=0.4]{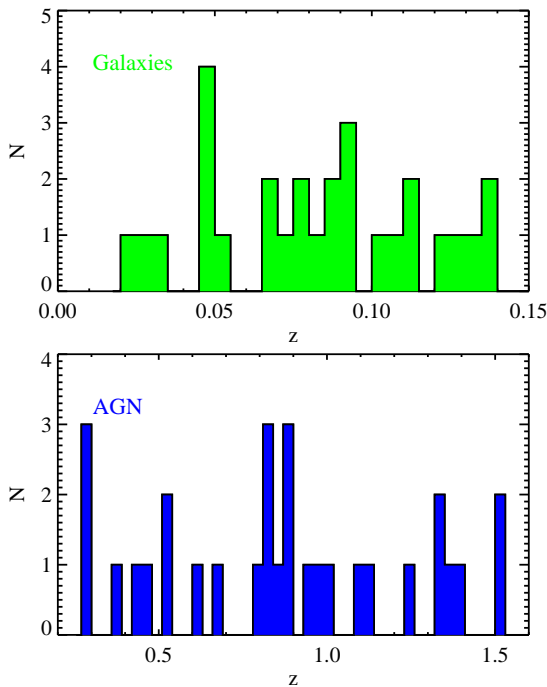}
\end{center}
\caption{Spectroscopic redshifts for the 60 extragalactic 
sources discovered in our survey.  The redshift range of
each panel has been truncated
for presentation purposes.
%The `spikes' in the histograms are likely the result
%of large-scale structures in the universe behind M31.
}
\label{fig:redshift}
\end{figure}

\begin{figure}
\begin{center}
\includegraphics[scale=0.3,angle=90]{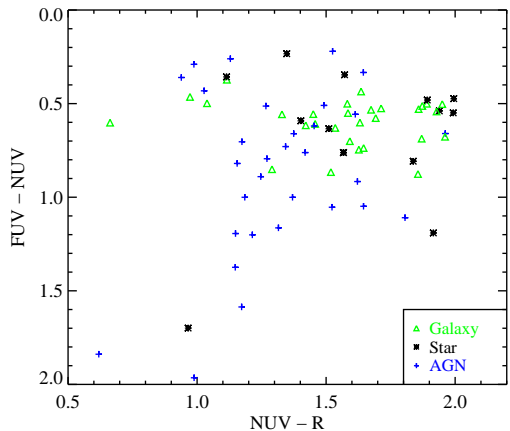}
\end{center}
\caption{Color-color plot of the 76 objects targeted for
follow-up optical spectroscopy.   The sources spectroscopically
classified as AGN show systematically redder $FUV-NUV$ color
and bluer $NUV-R$ than the galaxy and stars.
}
\label{fig:clrspec}
\end{figure}

Of the targets drawn from our primary candidate list for spectroscopic
study (59 total from
Table~\ref{tbl_allext_sub}),  85$\%$ were verified as 
extragalactic and half of these show broad-lines indicating strong AGN
activity.   The redshift distributions of the objects are given in 
Figure~\ref{fig:redshift}. 
%Even with this rather small sample, one can
%identify likely large-scale structures behind M31 (e.g.\ at $z=0.05, 0.82$).
The galaxies discovered by our survey
are generally blue, faint star-forming systems
with typical late-type spectra.  A significant number of the higher redshift
systems, however, have red continua and exhibit absorption lines characteristic
of early-type systems.  We interpret their significant UV fluxes as 
signatures of the so-called ``UV upturn'' observed in 
early-type galaxies \citep[e.g.][]{o99}.

Regarding the stellar contaminants, the majority of these were especially bright 
($R < 17$\,mag) and could have been avoided by imposing a brightness limit 
to the candidate criteria.
On the other hand, if any of these had been confirmed as AGN they would have
been especially valuable as probes of M31's halo.  
Figure~\ref{fig:clrspec} shows that the targets classified as AGN have 
systematically redder $FUV-NUV$ color and systematically bluer $NUV-R$ colors
than the typical galaxies observed.
By restricting the pre-selection to $FUV-NUV > 0.5$ and $NUV-R < 1.4$,
one would likely increase the efficiency of quasar selection.  The cut on
$FUV-NUV$ color, however, would also reduce the number of candidates with 
higher $FUV$ flux.  We plan for a future observing run in Fall 2009
to confirm additional candidates.

\begin{figure}
\begin{center}
\includegraphics[scale=0.4,angle=90]{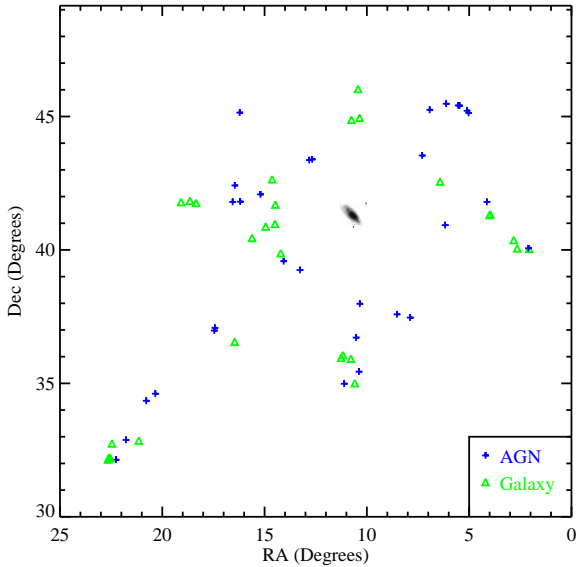}
\end{center}
\caption{Projected distribution of extragalactic sources in the
field surrounding M31 discovered or rediscovered (2 sources)
by our survey to date.  The AGN, which are optimal for follow-up
absorption-line spectroscopy of M31's halo, are distributed
through the projected halo with impact parameters ranging from
$\approx 35-180$\,kpc.
}
\label{fig:extragal}
\end{figure}

\section{Summary}
\label{sec:summ}

Adopting color-color criteria based on 
KPNO/$R$, GALEX/\nuv, and GALEX/$FUV$ photometry,
we have culled a list of $\approx 100$ quasar candidates 
(Table~\ref{tbl_allext_sub})
from 50 \sqdeg\ out to 150 kpc.  
We have obtained follow-up spectroscopy for 59 of these "tier 1" candidates
and an additional 27 objects with less restrictive color-color criteria.
We report the detection of 30 broad-lined targets which we classify as AGN
and one featureless source which is a previously known blazar.
We also report on 29 galaxies ranging from $z=0.02$ to $z=0.25$.

The AGN are distributed across M31's halo with impact parameters ranging
from 35 to 180\,kpc (Figure~\ref{fig:extragal}).  
Roughly half of these sources are sufficiently bright
for follow-up UV spectroscopy using the HST/COS spectrometer, albeit
with multi-orbit exposures.
Such spectra could be used to examine the spatial distribution of
enriched gas via absorption-line analysis of transitions like
$Si_{II}$~1260, $Si_{IV}$~1393,1402, and $C_{IV}$~1548,1550,
revealing the metallicity, gas kinematics, molecular content, and
the nature of M31's ionized gas.  As the identified QSOs are
distributed throughout M31, the sampling of this gas content
directly aids in our understanding of galaxy formation processes
through the deconvolution of disc/halo regions and the interpretation
of accretion events such as the giant southern stream.  
Recent work has characterized the properties and distribution of stars 
across these regions of M31, but an unambiguous separation of the
components requires a sampling of the gas dynamics.

Future surveys (e.g., PAndAS and PAndromeda) will soon lead to global
maps of M31 and we can target future QSO surveys to interesting
regions with/without substructure.  The area that is covered by 
our GALEX pointings is small relative to the (yet) unobserved
M31 halo through optical imaging.

Various surveys of M31, most notably the SDSS, have revealed many
QSO detections, but focused on the disc itself.  This project 
used those tools to extend the search to the newly discovered halo of M31.
Future searches for quasars toward M31 in pointings not covered by
this survey could provide an additional
hundred or more targets for such analysis.

\section*{Acknowledgments}

AF, JXP JSK, and PG acknowledge support NASA
grant NNG06GD37G, as part of the GALEX GO program.
PG and JSK acknowledge support from NSF grant AST-0607852.
J.~S.~was supported by NASA through a Hubble
Fellowship, administered by the Space Telescope Science Institute, which
is operated by the Association of Universities for Research in Astronomy,
Incorporated, under NASA contract NAS5-26555.

%\bibliographystyle{/u/xavier/paper/Bibli/mn2e}
%\bibliography{/u/xavier/paper/Bibli/allrefs}

%%%%%%%%%%%%%%%%%%%%%%%%%%
%%%%%%%%%%%%%%%%%%%%%%%%%% 

%\begin{figure}
%\begin{center}
%\scalebox{.4}{\includegraphics{../newM31_W2_2.opt.eps}}
%\end{center}
%\caption{To check KPNO's astrometry, we plot GALEX objects (blue crosses), 
%KPNO objects (green circles) and matches (green/blue circles). 
%The image shown is a small part of 'W2'.  GALEX detections with no KPNO
%matches are too faint to yield any useful photometry.  
%KPNO detections with no GALEX matches are very faint in the UV 
%and are likely red stars.  
%Image is 6 x 10 arcminutes, with north being left and east being up.  
%The star found by GALEX and KPNO close to the center is at 
%(RA,DEC) = (00:23:14.2, +42:34:23.5).}
%\label{w2}
%\end{figure}

%\begin{figure}
%\begin{center}
%\scalebox{.4}{\rotatebox{90}{\includegraphics{../sigmaplot.epsi}}}
%\end{center}
%\caption{Since we did not require extremely precise photometry, we calibrated 
%our R magnitudes against the USNO B1.0 catalog.  We plot the USNO and KPNO 
%magnitudes and calculated the offset between the linear functions of each set of data.  
%Points in red are more than 2 $\sigma$ from the mean and are eliminated.}
%\label{sigmaplot}
%\end{figure}

\end{document}